\documentclass[structabstract]{aa}  
\usepackage{graphicx}

\usepackage{subfigure}
\def\mean#1{{\langle}#1{\rangle}} 
\usepackage{txfonts}
\usepackage{natbib}
\bibpunct{(}{)}{;}{a}{}{,} 
\begin{document}

\title{A new instability domain of CNO-flashing low-mass He-core stars on their early white-dwarf cooling branches} 
\author{Leila M. Calcaferro\inst{1,2},
        Alejandro H. C\'orsico\inst{1,2},  
        Leandro G. Althaus\inst{1,2}, \and 
        Keaton J. Bell\inst{3,4}}
\institute{$^{1}$ Grupo de Evoluci\'on Estelar y Pulsaciones,
  Facultad de Ciencias Astron\'omicas y Geof\'isicas,
  Universidad Nacional de La Plata, Paseo del Bosque s/n,
  1900, La Plata, Argentina\\
  $^{2}$ Instituto de Astrof\'isica La Plata,
  CONICET-UNLP, Paseo del Bosque s/n, 1900, La Plata, Argentina\\
  $^{3}$ DIRAC Institute, Department of Astronomy, University of Washington, Seattle, WA-98195, USA\\
  $^{4}$ NSF Astronomy and Astrophysics Postdoctoral Fellow and DIRAC Fellow\\
\email{lcalcaferro@fcaglp.unlp.edu.ar}      }

\date{Received ;  }

\abstract {Before reaching their quiescent terminal white-dwarf cooling branch, some low-mass 
helium-core white dwarf stellar models experience a number of nuclear flashes which greatly reduce their hydrogen envelopes. Just before the occurrence of each flash, stable hydrogen burning may be able to drive global pulsations that could be relevant to shed some light on the internal structure of these stars through asteroseismology, similar to what happens with other classes of pulsating white dwarfs.}
  {We present a pulsational stability analysis applied to low-mass helium-core stars on their early white-dwarf cooling branches going through CNO flashes in order to study the possibility that the $\varepsilon$ mechanism is able to excite gravity-mode pulsations. We assess the ranges of unstable periods and the corresponding instability domain in the $\log g- T_{\rm eff}$ plane.} 
  { We carried out a nonadiabatic pulsation analysis for low-mass helium-core white-dwarf models with stellar masses between $0.2025$ and $0.3630\ M_{\sun}$ going through CNO flashes during their early cooling phases.}
  {We found that the $\varepsilon$ mechanism due to stable hydrogen burning can excite low-order ($\ell= 1, 2$) gravity modes with periods between $\sim 80$ and $500\ $s, for stars with $0.2025 \lesssim M_{\star}/M_{\sun} \lesssim 0.3630$ located in an extended region of the $\log g - T_{\rm eff}$ diagram with effective temperature and surface gravity in the ranges $15\,000 \lesssim T_{\rm eff} \lesssim 38\,000\ $K and $5.8 \lesssim \log g \lesssim 7.1$, respectively. For the sequences that experience multiple CNO flashes, we found that with every consecutive flash, the region of instability becomes wider, and the modes, more strongly excited. The magnitudes of the rate of period change for these modes are in the range $\sim 10^{-10}$ - $10^{-11}\ $[s/s].}
  {Since the timescales required for these modes to reach amplitudes large enough to be observable are shorter than their corresponding evolutionary timescales, the detection of pulsations in these stars is feasible. Given the current problems in distinguishing some stars that are populating the same region of the $\log g- T_{\rm eff}$ plane, the eventual detection of short-period pulsations may help in the classification of such stars.
 Furthermore, if a low-mass  white dwarf star were found to pulsate with low-order gravity modes in this region of instability, it would confirm our result that such pulsations can be driven by the $\varepsilon$ mechanism. In addition, confirming a rapid rate of period change in these pulsations would support that these stars actually experience CNO flashes, as predicted by evolutionary calculations.}  
  \authorrunning{Calcaferro et al.}
  \titlerunning{A new instability domain for CNO-flashing low-mass WDs}
  \maketitle
%

\section{Introduction}
\label{introduction}

White dwarf (WD) stars represent the end stage in the life of the 
majority of all stars, including our Sun \citep{2008ARA&A..46..157W,
2008PASP..120.1043F,2010A&ARv..18..471A,2019A&ARv..27....7C}. Most WDs 
\citep[$\sim 85\%$; see][]{2016MNRAS.455.3413K} are characterized
by hydrogen (H) atmospheres and are called DA WDs, having an average stellar mass of $\sim 0. 6\ M_{\sun}$ \citep{2019MNRAS.486.2169K}. According to the stellar evolution theory, they probably
harbor carbon-oxygen (CO) cores, although the most massive ones may 
have cores made of O and neon (Ne). At variance with average DA WDs, 
there is a population of WDs with low mass ($M_{\star} \lesssim 0.45\  M_{\sun}$), 
that probably harbor helium (He) cores. The formation context for such low-mass
WDs is thought to consist of a low-mass red giant-branch (RGB) star that 
experiences strong mass loss, mostly as a result of binary interaction taking place before the onset of the He flash \citep{2013A&A...557A..19A,2016A&A...595A..35I},
which is avoided, and the core of these stars is composed of He. The current evolutionary models predict that,
once the mass-loss stage is finished,  these low-mass stars would experience a 
number of CNO nuclear flashes which greatly reduce their H content
before reaching their quiescent terminal WD cooling branch. 
Such binary-star evolution scenario is confirmed by observations, since 
most low-mass WDs are found in binary systems \citep{1995MNRAS.275..828M}. 
Theoretical computations \citep[see, for instance,][]{2013A&A...557A..19A,2016A&A...595A..35I} predict  that
low-mass WDs with masses lower than $\sim 0.18- 0.20\  M_{\sun}$\footnote{This threshold depends on the metallicity of the WD progenitors \citep[e.g.][]{2002MNRAS.337.1091S,2016A&A...595A..35I}. Some authors adopt $\sim 0.3\  M_{\sun}$ as the upper mass limit for ELM WDs \citep{2016ApJ...818..155B}.}, called  extremely low-mass  (ELM) WDs,
do not experience CNO flashes in their past evolution. 
The absence of flashes would consequently suggest that ELM WDs harbor thick H envelopes, and then, they would be characterized by very long cooling timescales, and have pulsational properties different in comparison with systems that had experienced flashes \citep[see][]{2013A&A...557A..19A,2014A&A...569A.106C}.

In the last years, numerous low-mass and ELM WDs have been detected in the context of relevant surveys such as the SDSS, ELM, SPY and WASP  \citep[see, for instance,][]{2009A&A...505..441K,2010ApJ...723.1072B,2016ApJ...818..155B,2020ApJ...889...49B,2011ApJ...727....3K,2012ApJ...751..141K,2015ApJ...812..167G,2020ApJ...894...53K}.
The discovery of their probable precursors, the so-called low-mass pre-WDs, 
has triggered the interest in this type of objects because of the possibility of studying the evolution of the progenitors that lead to the WD phase. Even more interesting, the detection of multi-periodic brightness variations in low-mass WDs
\citep{2012ApJ...750L..28H,
  2013ApJ...765..102H,2013MNRAS.436.3573H, 2015MNRAS.446L..26K,2018MNRAS.479.1267K,
  2017ApJ...835..180B,2018A&A...617A...6B,2018MNRAS.478..867P}, and low-mass pre-WDs \citep{2013Natur.498..463M,2014MNRAS.444..208M,2016ApJ...822L..27G,
  2020ApJ...888...49W} 
has brought about  new classes of pulsating stars known as ELMVs and pre-ELMVs, respectively. It 
has allowed the study of their stellar interiors with the tools of asteroseismology,  similar to the case of other pulsating WDs such as ZZ Ceti stars or DAVs ---pulsating WDs with H-rich atmospheres--- and V777 Her or DBVs ---pulsating WDs with He-rich atmospheres \citep[][]{2008ARA&A..46..157W,2008PASP..120.1043F,2010A&ARv..18..471A,2019A&ARv..27....7C}. The pulsations observed in ELMVs are compatible with global gravity ($g$)-mode pulsations. In the case of pulsating ELM WDs, the pulsations have large 
amplitudes mainly at the core regions \citep{2010ApJ...718..441S,2012A&A...541A..42C,2014A&A...569A.106C}, allowing the study of their core chemical structure. According to nonadiabatic computations   \citep{2012A&A...541A..42C,2013ApJ...762...57V,2016A&A...585A...1C} these modes are probably excited by the $\kappa-\gamma$ \citep{1989nos..book.....U} mechanism, acting at the H-ionization zone. In the case of pre-ELMVs, the nonadiabatic stability computations for radial \citep{2013MNRAS.435..885J} and nonradial $p$- and $g$-mode pulsations \citep{2016A&A...588A..74C,2016ApJ...822L..27G,2016A&A...595L..12I} revealed that the excitation is probably due to the $\kappa-\gamma$ mechanism acting mainly at the zone of the second partial ionization of He, with a weaker contribution from the region of the first partial ionization of He and the partial ionization of H. 
The presence of He in the driving zone is crucial in order to have the modes destabilized by the $\kappa-\gamma$ mechanism \citep{2016A&A...585A...1C,2016A&A...595L..12I}.

Additionally, \cite{2014ApJ...793L..17C} showed that the $\varepsilon$ mechanism due to stable H burning may contribute to destabilize some short-period $g$ modes at the basis of the H envelope, in particular, for  low-mass WD sequences with stellar masses lower than $\lesssim 0.18 M_{\sun}$ and effective temperatures below $\sim 10\,000$ K. 
The $\varepsilon$ mechanism is thought to be a potential mechanism  for exciting pulsations in several type of stars. Since nuclear burning has a strong dependence on the temperature, it can lead to a pulsational instability induced by thermonuclear reactions. Indeed, this mechanism can be responsible for the the excitation of $g$ modes not only in ELM WDs \citep{2014ApJ...793L..17C}, but also in other types of WDs and pre-WDs as well. For instance, \cite{2009ApJ...701.1008C} showed that this mechanism could drive short-period $g$-mode pulsations in GW Virginis stars. In addition, \cite{10.1093/pasj/psu051}  found that this mechanism can excite low-degree $g$ modes 
in very hot DA WDs and pre-WDs coming from solar-metallicity progenitors.
Later, \cite{2015A&A...576A...9A} and \cite{2016A&A...595A..45C}  showed that DA WDs coming from low-metallicity progenitors can sustain stable H burning even at low luminosities, motivating a subsequent nonadiabatic exploration that demonstrates the excitation of low-order $g$ modes in hot DA WDs and pre-WDs \citep{calcaferro_2017_proceeding}, as well as in DA WDs with effective temperatures typical of ZZ Ceti stars \citep{2016A&A...595A..45C} coming from sub-solar metallicity progenitors. Finally,  attempts were done to explain the pulsations observed in other kind of compact stars, the pulsating He-rich hot subdwarf stars, as triggered by nuclear burning through the $\varepsilon$ mechanism \citep{2011ApJ...741L...3M,2018A&A...614A.136B}.

Another relevant aspect for this type of stars is the secular change of periods with time ($\dot{\Pi} $), which reflects the evolutionary timescale of these stellar remnants. In particular,
the theoretical computations of the rates of period change carried out by \cite{2017A&A...600A..73C} for the low-mass WD, pre-WD and pre-CNO flash stages, indicate that the magnitudes of $\dot{\Pi}$ of $g$ modes for models evolving in stages prior to the CNO flashes are up to $10^{-10}$ - $10^{-11}\ $[s/s], 
and more importantly, larger than the maximum magnitudes of $\dot{\Pi}$ predicted for the other two stages analyzed in that work (i.e., the WD and pre-WD stages). 

In this work, we present a nonadiabatic stability analysis considering the effects of the $\varepsilon$ mechanism in destabilizing $g$-mode periods for sequences of low-mass  WDs evolving through CNO flashes. The low-mass  WD sequences expected to experience CNO flashes are in the mass range of $0.2 \lesssim M_{\star}/M_{\sun} \lesssim 0.4$. \cite{2013A&A...557A..19A}  showed that, although these models evolve faster than their counterparts with masses below $0.18-0.20\ M_{\sun}$, it is possible to detect a star evolving prior to the flash where evolution clearly slows down \citep[see Fig. 4 of][]{2013A&A...557A..19A}.  A fundamental motivation in this paper is the possibility of finding $g$-mode pulsations in stars that go through the region in the $\log g-T_{\rm eff}$ diagram where the flashes take place. 
This will allow to study the structure of such objetcs, and consequently complement the information that can be extracted by analyzing the pulsations observed in ELMV and pre-ELMV stars \citep[e.g.,][]{2014A&A...569A.106C,2016A&A...585A...1C,2016A&A...588A..74C,2016A&A...595A..35I,2017A&A...600A..73C,2017A&A...607A..33C,2018A&A...620A.196C}.

We report the existence of a new instability strip for low-mass He-core  stars evolving through CNO flashes. We study the destabilization effects produced by the $\varepsilon$ mechanism due to stable H burning. The paper is organized as follows. 
A brief summary of the numerical codes and the stellar models employed is provided in
Sect. \ref{numerical}. In Sect. \ref{stability}, we show the results of the stability 
analysis performed on the sequences of low-mass  WDs under study.
Finally, in Sect. \ref{conclusions} we summarize the main
findings of this work. 

\section{Numerical codes}
\label{numerical}

We employed the evolutionary models of low-mass He-core WDs 
generated with the {\tt LPCODE} stellar evolution code \citep{2013A&A...557A..19A}. 
{\tt LPCODE} evolutionary code computes the complete evolutionary 
stages that lead to the formation of the WD, thus
allowing the study of the WD evolution consistently with the predictions of the
evolutionary history of the progenitors. Initial configurations
for low-mass He-core WD models were computed by \cite{2013A&A...557A..19A}
by mimicking the binary evolution   of an initially $1.0\ M_{\sun}$ solar 
metallicity donor star and a $1.4\ M_{\sun}$ neutron star companion. 
Binary evolution was assumed to be fully nonconservative,
and the losses of angular momentum due to mass loss, gravitational wave
radiation, and magnetic braking were considered. 
Initial He-core WD models with stellar masses between $0.1554$ and $0.4352\ M_{\sun}$
characterized by thick H envelopes were derived from stable mass loss
via Roche-lobe overflow, see \cite{2013A&A...557A..19A} for details. During
WD regime, time-dependent element diffusion due to gravitational settling and chemical and thermal 
diffusion of nuclear species was considered, following the multicomponent
gas treatment of \cite{1969fecg.book.....B}. 
For this work, we analyzed the sequences with stellar masses 
of $0.2025$, $0.2724$, $0.3207$ and $0.3630M_{\sun}$. 
These sequences were evolved through the stages of multiple thermonuclear CNO 
flashes that take place during their early cooling branch.

We carried out a pulsation stability analysis of nonradial dipole ($\ell= 1$) and quadrupole ($\ell= 2$) $g$ modes employing the nonadiabatic  version of the {\tt LP-PUL} pulsation code \citep[see,][for details]{2006A&A...458..259C,2009ApJ...701.1008C}. This pulsation code 
is  based  on  a  general Newton-Raphson technique that solves the sixth-order complex system of linearized equations and boundary conditions 
\citep[see][]{1989nos..book.....U}. The Brunt-V\"ais\"al\"a
frequency ($N$) was computed following the so-called ``Ledoux Modified''
treatment \citep{1990ApJS...72..335T,1991ApJ...367..601B}. Our nonadiabatic 
computations are based on the frozen-convection approximation, that 
neglects the perturbation of the convective flux. 
The set of pulsation modes considered in this work covers a very wide range of periods (up to $\sim 6000\ $s), in order to properly determine the upper limits of the instability domain. We have also carried  out additional calculations neglecting the effects of nuclear energy release on the nonadiabatic pulsations, for which we set $\varepsilon= \varepsilon_{\rho}= \varepsilon_T= 0$, being $\varepsilon$ the nuclear energy production rate, and $\varepsilon_{\rho}$ and $\varepsilon_T$ the corresponding logarithmic derivatives $\varepsilon_{\rho}= (\partial{\ln \varepsilon}/\partial{\ln \rho})_T$ and $\varepsilon_T= (\partial{\ln \varepsilon}/\partial{\ln T})_{\rho}$. This prevents the $\varepsilon$ mechanism from operating, however nuclear burning was still taken into account in the evolutionary calculations.

\begin{figure} 
\begin{center}
\includegraphics[clip,width=9 cm]{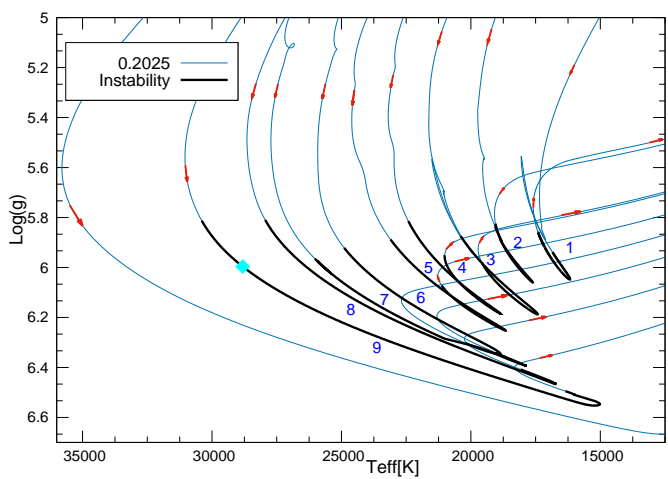} 
\caption{Instability domain corresponding to the $\varepsilon$ mechanism on 
the $\log g$ vs $T_{\rm eff}$ diagram for the low-mass WD  sequence with $0.2025\ M_{\sun}$.  The stages of pulsational instability are emphasized with thick black lines along the evolutionary tracks. Low-order $g$ modes are driven before each flash. The track begins after the end of Roche-lobe overflow (upper right branch of the curve) and proceeds downward, toward higher values of $T_{\rm eff}$, until the first CNO flash takes place. Numbers denote every consecutive flash. Red arrows 
along the curve indicate the course of the evolution. The cyan diamond  before the ninth flash indicates the location of the template model analyzed in Fig. \ref{dTT_dwdr_w}.} 
\label{HR_0202} 
\end{center}
\end{figure}

\section{Stability analysis}
\label{stability}

\begin{figure*} 
\begin{center}
\includegraphics[clip,width=19.1 cm]{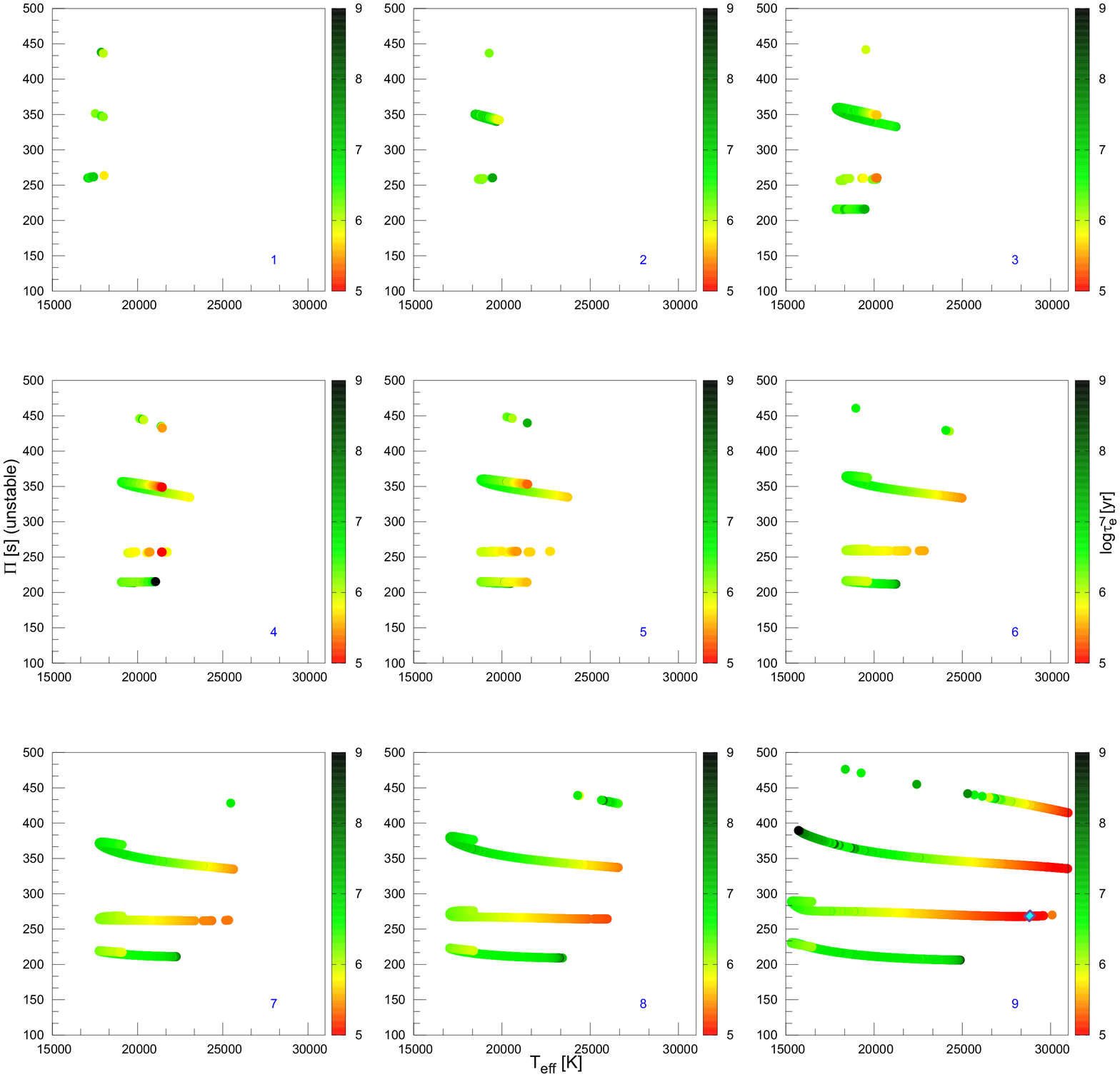} 
\caption{Periods of unstable $\ell= 1$ $g$ modes in terms of the effective temperature for the nine flashes experimented by our template low-mass  sequence with $0.2025\ M_{\sun}$. Color coding indicates the logarithm of the $e$-folding time ($\tau_e$) of each unstable mode (right scale). Blue numbers 
at the bottom right corner of each panel indicate the number of the 
flash, as in Fig. \ref{HR_0202}. The cyan diamond on the bottom right panel indicates the location of the template model analyzed in Fig. \ref{dTT_dwdr_w} (as in Fig. \ref{HR_0202}).}
\label{instability_0202} 
\end{center}
\end{figure*}

\begin{figure} 
\begin{center}
\includegraphics[clip,width=9.0 cm]{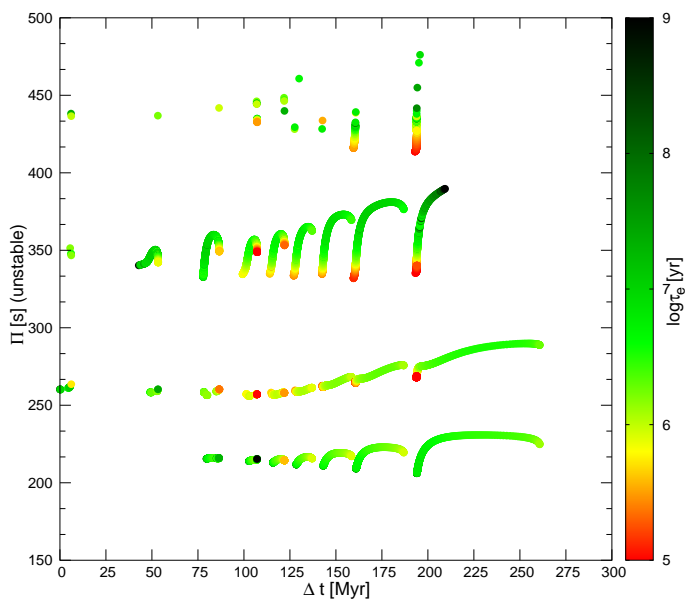} 
\caption{Same as Fig.\ref{instability_0202} for the nine flashes of the sequence with $0.2025\ M_{\sun}$, but in terms of the elapsed time since the appearance of the first unstable mode for the first flash.}
\label{instability_age_0202} 
\end{center}
\end{figure}

\begin{figure}
\begin{center}
\includegraphics[clip,width=8 cm]{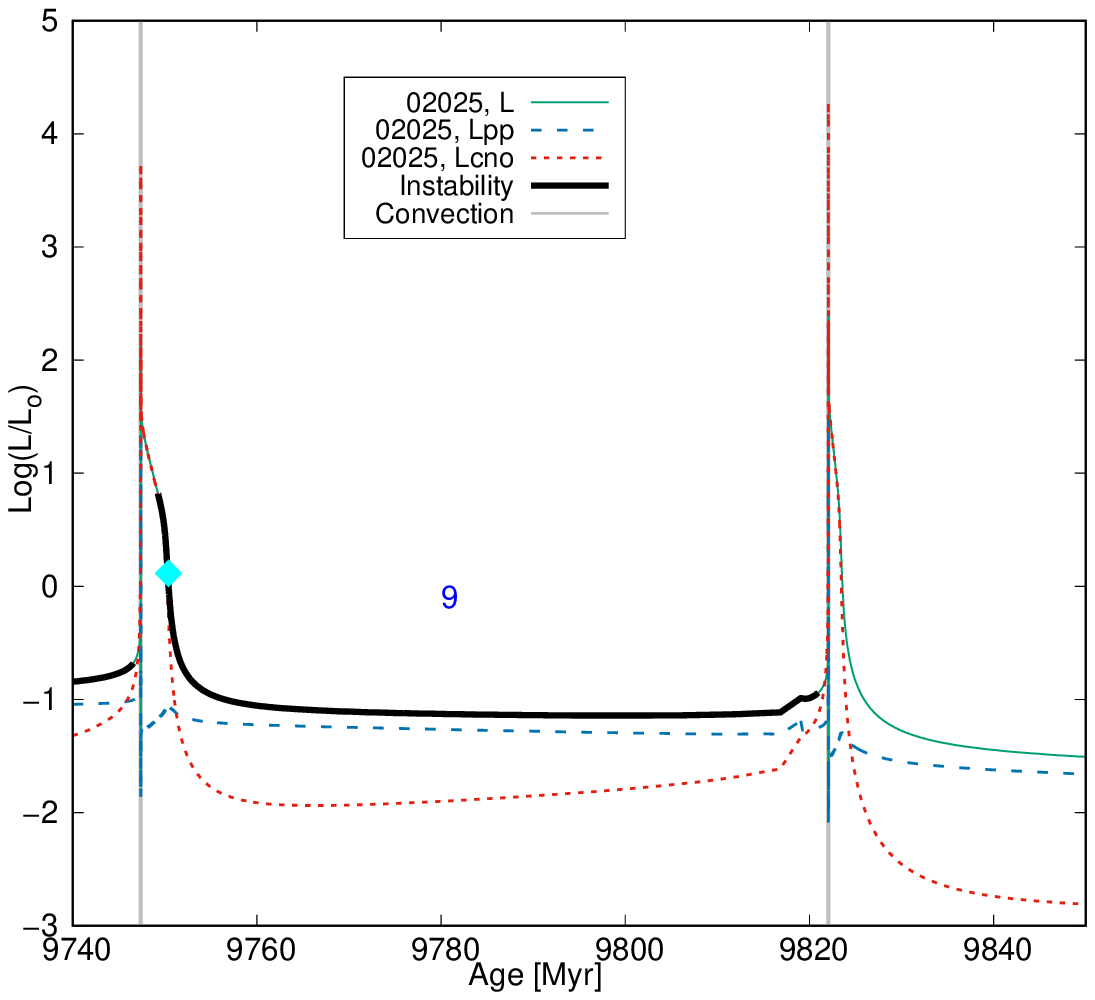}
\caption{$\log(L/L_{\sun})$ vs age (in Myr) for the sequence with $0.2025\ M_{\sun}$, corresponding to the last instability region. The temporal evolution of the surface luminosity (green line), the luminosity given by the pp chain (light-blue dashed line) and the luminosity due to the CNO bicycle (red dotted line) are shown, along with the ninth instability region emphasized by a thick black line. Grey vertical strips mark the narrow regions where convection is present (either internal or external). A blue number indicates the number of the flash (as in Fig. \ref{HR_0202}), while the cyan diamond indicates the location of the template model analyzed in Fig. \ref{dTT_dwdr_w} (and \ref{HR_0202}).}
\label{lum_age}
\end{center}
\end{figure}

To illustrate the results of our nonadiabatic study, we display in Fig.~\ref{HR_0202} the evolutionary track of the low-mass He-core  sequence with $0.2025\ M_{\sun}$ on the $\log g$ vs $T_{\rm eff}$ plane. As shown by \cite{2013A&A...557A..19A}, this sequence experiences nine CNO flashes before entering the final cooling track, that reduce considerably the thickness of the H envelope. Before and along each loop described by this evolutionary sequence, our nonadiabatic exploration shows that the $\varepsilon$ mechanism is able to destabilize  low-order $\ell= 1$ $g$ modes. We highlight the corresponding region of instability of each part of the track with thick black lines. It is clear that the extension of the region of instability in the $\log g$ vs $T_{\rm eff}$ plane grows with every consecutive flash. The periods of unstable $\ell= 1$ $g$ modes in terms of the effective temperature for each one of the loops described by this sequence  are shown in Fig.~\ref{instability_0202}. In the panels, the color coding indicates the logarithm of the $e$-folding time (in years) of the unstable modes, which represents a measure of the time taken for the perturbation that causes the oscillation to reach an observable amplitude. Its definition is given by the expression $\tau_e = 1/|\Im(\sigma)|$, where $\Im(\sigma)$ is the imaginary part of the complex eigenfrequency  $\sigma$. For the panel representing the first CNO flash (top left panel), marked as "1", there are only a few unstable $\ell= 1$ $g$ modes with low radial orders ($k= 2$, $3$ and $4$) corresponding to periods between $\sim 260$ and $450\ $s, in a very limited range of effective temperature ($\sim 16\,000 - 18\,000\ $K), and with relatively large values for the $e$-folding time (being its minimum value $\sim 5.4 \times 10^5\ $yr). The Figure shows that with every consecutive flash, the instability region becomes wider, and with more modes being destabilized. For instance, in the third flash (top right panel), one additional mode is excited, corresponding to $k= 1$, and in general, the values of the $e$-folding time shorten as indicated by the color coding. It is apparent that for the last three panels ("7", "8" and "9") the instability domains are considerably extended and the $e$-folding times significantly shorten, that is, the excitation becomes stronger, with the implication that these modes may have a larger chance to reach observable amplitudes. For the ninth panel (bottom right panel), it is evident the large region in the diagram where we can find $\ell= 1$ $g$-mode periods destabilized by the $\varepsilon$ mechanism. In this case, the periods are characterized by 
$k= 1$, $2$, $3$ and $4$ corresponding to periods in the range of $\sim 200$ and $480\ $s, effective temperatures between $\sim 15\,000$ and $31\,000\ $K, and with $e$-folding times that, in its lowest, reach down to $\sim 8.2 \times 10^4\ $yr. 
We note  that, in general, there is a trend for the periods to lengthen as the evolution proceeds to lower effective temperatures, before entering the loop, and to shorten afterward.

\begin{figure*} 
\begin{center}
\subfigure{\label{fig:compar_1}\includegraphics[clip,width=9.1 cm]{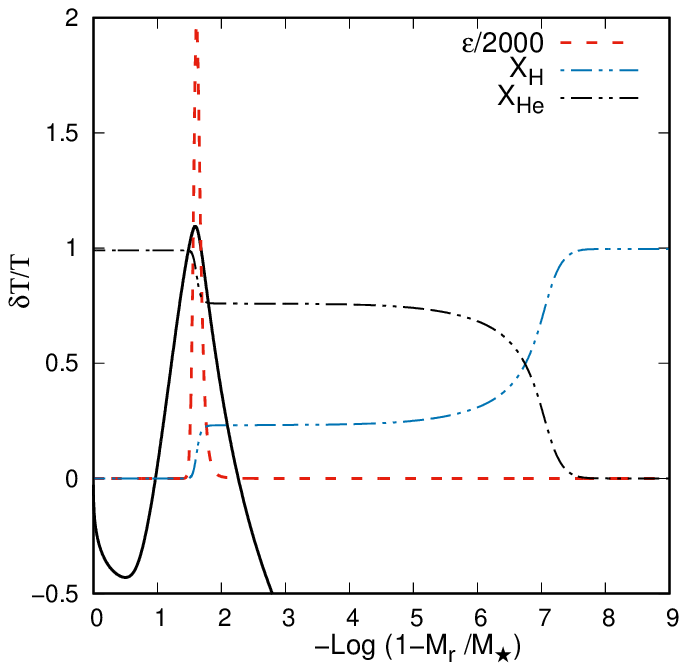}}
\subfigure{\label{fig:compar_2}\includegraphics[clip,width=9.1 cm]{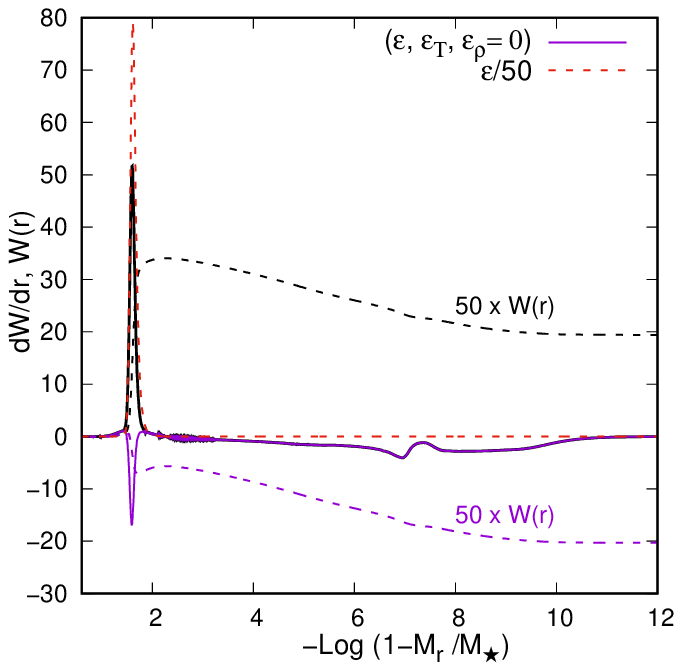}}
\caption{\textit{Left panel:} Lagrangian perturbation of the temperature ($\delta T/T$), along with the scaled nuclear generation rate ($\varepsilon$), and the fractional 
abundances of H and He ($X_{\rm H}$ and $X_{\rm He}$) in terms of the mass fraction coordinate ($-\log (1- M_r/M_{\star})$), for a representative unstable pulsation mode with
$\ell= 1$, $k= 2$ ($\Pi= 268.7\ $s), corresponding to a template model with $T_{\rm eff}= 28814\ $K of the $0.2025\ M_{\sun}$ sequence before the occurrence of the ninth CNO flash (see  Figs. \ref{HR_0202} and \ref{instability_0202}).  \textit{Right panel:} 
differential work function ($dW(r)/dr$) in terms of $-\log (1- M_r/M_{\star})$ for 
the case in which the $\varepsilon$ mechanism is allowed to operate (solid black curve) and
when it is suppressed (solid violet curve). Also shown are the corresponding 
scaled running work integrals, $W$ (dashed curves).}
\label{dTT_dwdr_w} 
\end{center}
\end{figure*}

In order to estimate if it would be possible to observe a star pulsating by the $\varepsilon$ mechanism while evolving through one of these loops, we consider the example of the models evolving through the ninth flash for the $0.2025\ M_{\sun}$ sequence. Given that the time spent by this sequence in the region of unstable $g$-mode periods for the ninth loop is $\sim 7.2 \times 10^6\ $yr, and the values of the $e$-folding times for many modes are significantly lower, $\sim 8.2 \times 10^4\ $yr, then these modes would have enough time to reach observable amplitudes. Since the duration of the whole pre-WD stage for this sequence, throughout the flashes and until the sequence gets to the maximum effective temperature, is $5.11 \times 10^8\ $yr, when we compare it to the time spent by the sequence during this ninth stage of instability, we see that it might be possible to detect a star with $0.2025\ M_{\sun}$ pulsating in low-order $g$ modes via $\varepsilon$ mechanism while evolving during this stage. Such possibility considerably increases if we take into account all the nine stages of instability experienced by this sequence. 
This can be visualized by showing how these pulsations vary with time. In Fig.~\ref{instability_age_0202} we display the periods of unstable $\ell= 1$ $g$ modes as in Fig.~\ref{instability_0202}, but in terms of the elapsed time (in Myr) since the appearance of the first unstable mode (corresponding to the first flash) for this sequence. It is clear that the flashes occur sooner each time, leading to less (and even almost negligible) temporal gaps with every consecutive flash. In addition, in Fig.~\ref{lum_age} we show the temporal evolution of the surface luminosity, $L$ (green line), the luminosity given by the pp chain, $L_{\rm pp}$ (light-blue dashed line), and the luminosity due to the CNO bicycle, $L_{\rm CNO}$ (red  dotted line), along with a thick black line emphasizing, particularly, the location of the models within the ninth instability region (as in Fig. \ref{HR_0202}). In the Figure, two very narrow grey vertical strips represent the evolutionary stages where convection (either internal or external) is present. 
The Figure shows that the region of instability starts when $L$ and $L_{\rm CNO}$ considerably drop after the occurrence of the eighth CNO flash and ends before the beginning of the ninth CNO flash. It is clear that during the flashes (and very shortly before, although not noticeable by its narrowness), an internal induced-flash convective zone develops, that quickly moves toward the stellar surface to 
rapidly vanish. 
In summary, at the stages of pulsation instability driven by the $\varepsilon$ mechanism, there is no convection inside our models. We conclude that convection does not affect any of the instability regions presented in this work.

Additionally, our calculations show that the $\varepsilon$ mechanism is also able to destabilize low-order $\ell= 2$ $g$ modes in all the sequences analyzed. Although the ranges of effective temperature of the models in which these modes are destabilized are approximately the same as in the $\ell= 1$ case, we found that the $\varepsilon$ mechanism destabilizes a higher number of $\ell= 1$ modes.
For instance, while for the ninth flash of the template sequence with $0.2025\ M_{\sun}$ we found in the $\ell= 1$ case that the modes with radial order $k$ from $1$ to $4$ are excited, in the $\ell= 2$ case, the modes with $k= 4$ remain stable. Also, the minimum value of the $e$-folding time is, in general, lower for the $\ell= 1$ case.
As expected, the range of unstable periods for $\ell= 2$ $g$ modes is shifted toward shorter values when compared to the ones found for the $\ell = 1$ case. In general, we found that the range of unstable periods for $\ell= 2$ $g$ modes spans from $\sim 80$ to $\sim 250\ $s. For brevity, in what follows we will focus on the $\ell= 1$ case.

\begin{figure*} 
\begin{center}
\includegraphics[clip,width=18 cm]{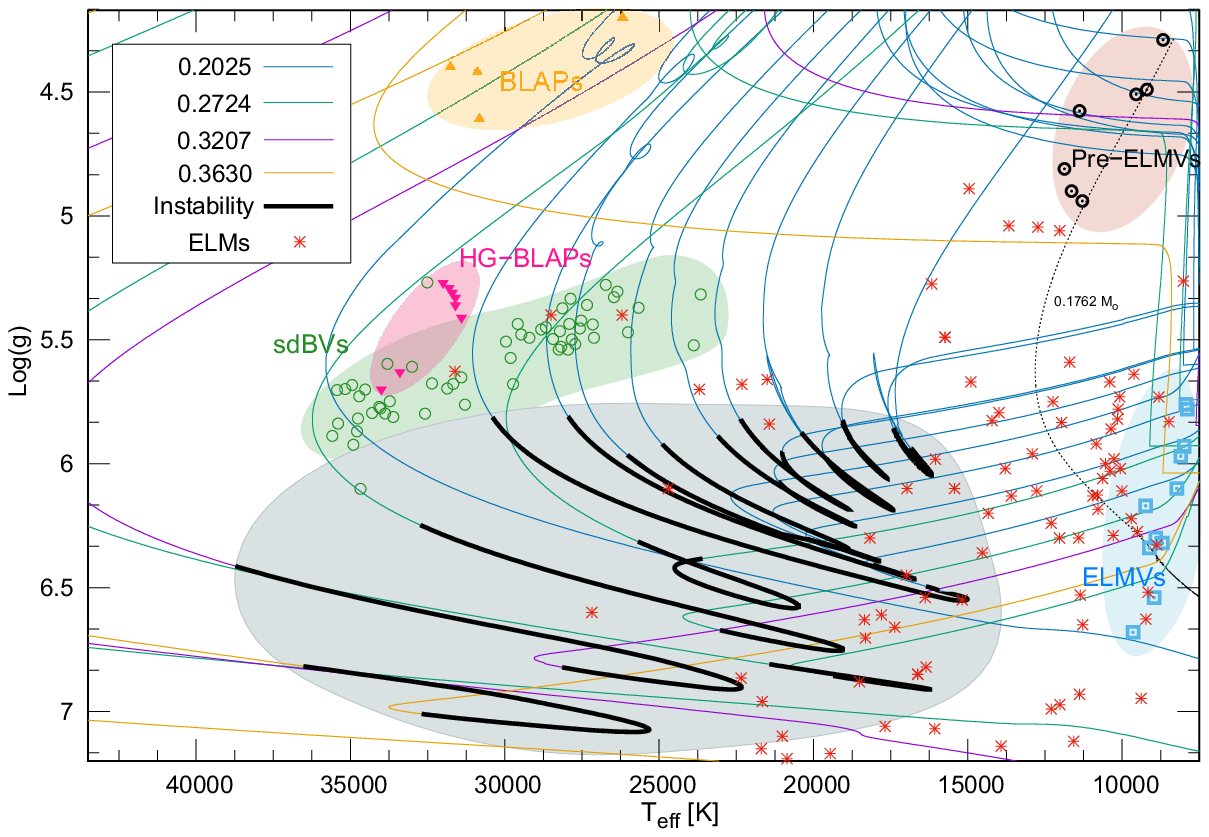}
\caption{Instability domain (grey shaded area) of low-order $g$ modes excited by the $\varepsilon$ mechanism on the $\log(g)$ vs $T_{\rm eff}$ for each sequence of low-mass  WD analyzed in this work. Stellar sequences are in units of solar mass. Red asterisks represent ELM (and pre-ELM) stars \citep{2010ApJ...723.1072B,2013ApJ...769...66B,2016ApJ...818..155B,2020ApJ...889...49B, 2011ApJ...737L..16V,2015ApJ...812..167G,2015MNRAS.450.3514K,2019MNRAS.488.2892P}. Light-blue squares with dots represent pulsating low-mass WDs (ELMVs) \citep{2012ApJ...750L..28H,2013ApJ...765..102H,2013MNRAS.436.3573H,2015MNRAS.446L..26K,2018MNRAS.479.1267K,2017ApJ...835..180B,2018A&A...617A...6B,2018MNRAS.478..867P}, while black circles with dots correspond to pulsating low-mass pre-WDs (pre-ELMVs) \citep{2013Natur.498..463M,2014MNRAS.444..208M,2016ApJ...822L..27G,2020ApJ...888...49W}. Green circles correspond to sdBVs \citep{2011ApJ...734...59G}, orange triangles represent BLAPs \citep{2017NatAs...1E.166P}, and pink triangles correspond to HG (High Gravity)-BLAPs \citep{2019ApJ...878L..35K}. The evolutionary track of the He-core WD sequence with $0.1762\ M_{\sun}$ is included as a reference.}
\label{instability_HR}
\end{center}
\end{figure*}

In order to show the role of the $\varepsilon$ mechanism as a destabilizing agent in CNO-flashing low-mass  WDs on their early-cooling branches, we pick out  a 
representative unstable pulsation mode corresponding to a template model 
evolving at the stage before the ninth flash for the sequence with $0.2025\ M_{\sun}$, indicated in Figs. \ref{HR_0202} and \ref{instability_0202} with a cyan diamond.
In the left panel of Fig.~\ref{dTT_dwdr_w}, we display the Lagrangian perturbation of the temperature, ($\delta T/T$), in terms of the mass fraction coordinate [$-\log (1- M_r/M_{\star})$], for the unstable $\ell= 1$, $k= 2$ $g$ mode ($\Pi= 268.7\ $s). Also shown are the scaled nuclear generation rate ($\varepsilon$), and the fractional abundances of H and He ($X_{\rm H}$ and $X_{\rm He}$).  The eigenfunction $\delta T/T$ has its maximum value at $-\log (1- M_r/M_{\star}) \sim 1.6$, where the H-burning shell is located. The $\varepsilon$ mechanism yields substantial driving to those $g$ modes that have their maximum of $\delta T/T$ at the narrow region of the burning shell \citep{1986ApJ...302..530K}.  It is illustrated for the $\ell =1$, $k= 2$
representative mode in the right panel of Fig.~\ref{dTT_dwdr_w}, in which we display 
with solid black curves the differential work function ($dW/dr$) in terms of 
$-\log (1- M_r/M_{\star})$, and also the scaled running work integral ($W$) 
with dashed black curves. It is evident that there is considerable driving 
($dW/dr > 0$) at the region of the H-burning shell for the mode analyzed. This mode
is globally unstable, as indicated by the positive value of $W$ at the stellar surface [$-\log (1- M_r/M_{\star})\sim 12$]. We have also 
performed additional stability computations in which we suppress the action of the $\varepsilon$ mechanism by forcing $\varepsilon= \varepsilon_{\rho}= \varepsilon_T= 0$ in the pulsation equations (see Sect. ~\ref{numerical}). The results for 
$dW/dr$ are shown in the right panel of Fig.~\ref{dTT_dwdr_w} with solid violet curves. Note that, in this case,  a strong damping ($dW/dr < 0$) takes place in the burning-shell region, resulting in a value $W < 0$ at the surface, which indicates that the mode is pulsationally stable. We conclude that, for this selected template model, the $k= 2$ $g$ mode is 
unstable due to the destabilizing effect of the H-burning shell via the $\varepsilon$ mechanism.

In Fig.~\ref{instability_HR} we show the evolutionary tracks of the low-mass He-core  WDs with $0.2025$, $0.2724$, $0.3207$ and $0.3630\ M_{\sun}$ on the $\log g$ vs $T_{\rm eff}$ plane. As in Fig. \ref{HR_0202}, thick black lines superimposed on every evolutionary track represent the regions where the $\varepsilon$ mechanism is able to excite low-order  $g$ modes. We have included a sample of ELM  WD stars \citep[shown with red asterisks;][]{2010ApJ...723.1072B,2013ApJ...769...66B,2016ApJ...818..155B,2020ApJ...889...49B, 2011ApJ...737L..16V,2015ApJ...812..167G,2015MNRAS.450.3514K,2019MNRAS.488.2892P}.
Also, we have included the location of the known ELMVs \citep{2012ApJ...750L..28H,
  2013ApJ...765..102H,2013MNRAS.436.3573H, 2015MNRAS.446L..26K,2018MNRAS.479.1267K,
  2017ApJ...835..180B,2018A&A...617A...6B,2018MNRAS.478..867P} marked with light-blue squares with dots, and pre-ELMVs \citep{2013Natur.498..463M,2014MNRAS.444..208M,2016ApJ...822L..27G,2020ApJ...888...49W}, indicated with black circles with dots (both regions have been emphasized with light-blue and light-red shaded areas, respectively). In addition, we have included the location of some sdBV stars \citep{2011ApJ...734...59G}, BLAPs \citep{2017NatAs...1E.166P}, as well as High-Gravity (HG)-BLAPs \citep{2019ApJ...878L..35K} indicated with green circles, and orange and pink triangles, respectively (their regions emphasized with green, orange and pink shaded areas, respectively). 
The Figure shows that every evolutionary sequence considered in this work has an extended zone of pulsation instability which, altogether, results in a wide region in the $\log g$ vs $T_{\rm eff}$ plane (grey shaded area) where low-order $\ell= 1$ $g$ modes can be destabilized by the $\varepsilon$ mechanism. This region covers the approximate ranges in $T_{\rm eff}$  and $\log g$ of $[15\,000 - 38\,000]$ K and $[5.8 - 7.1]$, respectively.  It is clear from the Figure that this new domain of instability does not overlap with the domain of instability of ELMVs, that lies at $T_{\rm eff} \lesssim 10\,000\ $K and similar values of $\log g$, nor with the one of the pre-ELMVs, that correspond to $T_{\rm eff} \lesssim 12\,000\ $K and $\log g \lesssim 5$. Note that the low-gravity boundary of the instability domain reported in this paper slightly overlaps with the high-gravity limit of the instability domain of the sdBV stars.

\begin{table*}
\centering
\caption{Stellar mass, total evolutionary timescale, time spent during the excitation phase, radial order, approximate range of effective temperature for the instability, average period, minimum value of the $e$-folding time of unstable $\ell= 1$ $g$ mode destabilized by the $\varepsilon$ mechanism.}
\begin{tabular}{ccccccc}
\hline
\hline
$M_{\star} [M_{\sun}]$ & $\Delta t_{\rm evol}$ [$10^6 $yr] & $\Delta t_{\rm exc}$ [$10^6 $yr] & $k$&  $T_{\rm eff}$ [kK]& $\mean{\Pi}$ [s] & $\tau_{e}$ [$10^6 $yr]\\
\hline
$0.2025$& $511$ & $160$ &  $1$&   $15$ - $25$   & $215$ & $0.380 $  \\ 
    $ $ & $ $   & &   $2$&   $15$ - $30$   & $259$ & $0.093 $  \\ 
    $ $ & $ $   & &   $3$&   $16$ - $31$   & $351$ & $0.081 $  \\
    $ $ & $ $   & &   $4$&   $18$ - $31$   & $442$ & $0.079 $  \\
   \hline
$0.2724$ & $135$& $108$  &  $1$&   $16$ - $30$   & $177 $ & $0.042 $  \\ 
     $ $ & $ $ &    &  $2$&   $16$ - $35$   & $205$ & $0.014$  \\
     $ $ & $ $ &    &  $3$&   $20$ - $36$   & $263 $ & $0.065$  \\
\hline    
$0.3207$ &  $18.0$& $10.7$  &  $1$&   $23$ - $29$   & $147$ & $0.068$ \\
     $ $ &  $ $ &   &  $2$&        $23$ - $35$   & $166$ & $0.160$   \\
     $ $ &  $ $ &   &  $3$&        $28$ - $38$   & $217$ & $0.240$   \\
\hline
$0.3630$ & $15.5$& $10.6$   &  $1$&   $26$ - $33$   & $130$  & $0.043$ \\
     $ $ &    $ $&     $ $  &  $2$&   $26$ - $36$   & $151$  & $0.100$ \\
     $ $ &    $ $&     $ $  &  $3$&   $30$ - $36$   & $191$  & $1.380$   \\
\hline 
\end{tabular}
\label{table_01}
\end{table*}

We summarize in Table~\ref{table_01} the main results of the stability analysis carried out for all the evolutionary sequences considered in this work. In the Table, the second column shows the total evolutionary timescale, $\Delta t_{\rm evol}$, which represents the time spent by the sequence between the beginning of the pre-WD phase and the maximum effective temperature reached before entering the final cooling branch. The third column indicates the time interval ($\Delta t_{\rm exc}$) in which stellar models exhibit pulsation instability due to the $\varepsilon$ mechanism, where we have considered the summation of the time taken by every instability phase for those sequences that experience multiple CNO flashes (that is, the total time that the models of a given sequence spend while evolving along the black line marked on its evolutionary track as indicated in Fig. \ref{instability_HR}).
The rest of the columns represent the radial order, the range of effective temperature of the instability, the average value of the periods, and the minimum value of the $e$-folding time of the unstable $\ell= 1$ $g$ modes destabilized by the $\varepsilon$ mechanism, respectively (where we have taken all the flashes into account for the multiple-flashing sequences).
Comparing the values of $\Delta t_{\rm exc}$ to the $e$-folding times, we see that for all the sequences there is plenty of time for the instabilities to reach observable amplitudes, being the unstable modes of the sequences with $0.2025$ and $0.2724\ M_{\sun}$ the ones more likely to be observed due to the larger differences between $\tau_{e}$ and $\Delta t_{\rm exc}$. When we compare the values of the total evolutionary timescale $\Delta t_{\rm evol}$ to $\Delta t_{\rm exc}$, we find that it would be possible to detect one of these low-mass  WD stars while evolving along these stages of instability where the $\varepsilon$ mechanism excites low-order $g$ modes.

It is interesting to estimate how many stars pulsating by the $\varepsilon$ mechanism are expected to be found. According to Table \ref{table_01}, considering once again the total time spent by the models of every evolutionary sequence during their stages of instability with respect to the evolutionary timescale, we can roughly estimate that between $\sim 31- 80 \%$ of these low-mass  WD stars may be found pulsating due the $\varepsilon$ mechanism. Considering the samples shown in Fig. \ref{instability_HR}, by virtue of their spectroscopic parameters, and taking the minimum value for the estimated probability, it might indicate that roughly $7$ of the stars in  the  catalogues could be found pulsating in low-order $g$ modes by the $\varepsilon$ mechanism. One of the reasons why these short-period pulsations have not been detected yet may be possibly attributed to the pulsation amplitudes being smaller than the detection limits.

In addition, the $\varepsilon$ mechanism is also able to destabilize a wide range of $g$-mode periods during the stages of the evolution between flashes, for all the sequences analyzed in this work. However, the $e$-folding times are larger than (or of the order of) the corresponding evolutionary timescales, and also, these evolutionary stages occur faster than the rest of the stages of the evolution \citep[e.g., $\sim 5.5 \times 10^3\ $yr for the sequence with $0.3630\ M_{\sun}$; see also][]{2013A&A...557A..19A}, so it would be unlikely to detect a pulsating star while evolving between flashes, 
let alone, exhibiting detectable pulsations. Then, we have discarded such regions as 
possible locations for stars pulsating by this mechanism.

We close this section by noting that we have carried out additional nonadiabatic calculations on model sequences with stellar masses $\lesssim 0.18-0.20\ M_{\sun}$, which, as we already mentioned, do not experience CNO flashes and  evolve very slowly. This was done in order to explore if in some part of the pre-WD evolution, the $\varepsilon$ mechanism is able to drive pulsations in these sequences.  In stages previous to the maximum effective temperature, H-shell burning via CNO bicycle is the dominant nuclear source for these sequences. We have found that the $\varepsilon$ mechanism is not capable of destabilizing $g$ modes for sequences with $M_{\star} \lesssim 0.18-0.20\ M_{\sun}$.

For the sake of completeness, we have also performed additional calculations for sequences with $M_{\star} \gtrsim 0.18-0.20\ M_{\sun}$, but this time artificially disabling the action of the element diffusion in the evolution of our stellar models, which has the effect of suppressing (or diminishing) the occurrence of the CNO flashes. As a consequence, for instance, the sequence with $0.2025\ M_{\sun}$  does not experience any CNO flashes, in agreement with the literature \citep[e.g.,][]{1998A&A...339..123D,2016A&A...595A..35I}. We have found that, although the $\varepsilon$ mechanism continues to destabilize some (but significantly less) $g$-mode period pulsations due to residual H burning, the corresponding values of the rate of period change of $g$ modes ($\sim 10^{-14}$ - $10^{-16}\ $[s/s]) are orders of magnitude lower than for modes of stellar models that go through flashes (as mentioned, $\sim 10^{-10}$ - $10^{-11}\ $[s/s]). Therefore, if one low-mass  WD was found to pulsate in this region of the $\log g$- $T_{\rm eff}$ diagram, and the rate of period change could be measured, it would help in discerning whether or not the star experiences CNO flashes. At the same time, since the values of the rate of period change  are much  lower for the non-flashing sequences, it would be very difficult to detect those.
 However, if a rate of period change was measured and resulted in a value lower than $\sim 10^{-12}\ $[s/s] then, based on our results, it would be possible to rule out that such a star is going through a flashing cycle.

\section{Summary and conclusions}
\label{conclusions}

In this paper, we  performed a stability analysis focused on low-mass  WD stars evolving through CNO flashes. We have shown that the $\varepsilon$ mechanism due to stable H burning is able to destabilize some low-order $\ell= 1, 2$ $g$ modes in stellar models with masses in the range of $0.2025 - 0.3630\ M_{\sun}$. As displayed in Figs. \ref{HR_0202} and \ref{instability_0202} for the template sequence with $0.2025\ M_{\sun}$, the sequences have more modes destabilized in every consecutive flash, and with shorter $e$-folding times. For several modes, the $e$-folding times are shorter than the corresponding evolutionary timescales, and therefore, there would be enough time to excite such pulsations to reach observable amplitudes. In general, this is true for all the sequences studied (see Table \ref{table_01}). The instability domain found is located in the ranges of $T_{\rm eff} \sim 15\,000 - 38\,000\ $K and $\log g \sim 5.8 - 7.1$ (see Fig. \ref{instability_HR}), and therefore, it does not overlap with the 
already known domains of instability of ELMVs and pre-ELMVs, and  
barely  overlaps ---at its low-gravity boundary--- with the the instability domain of the sdBV stars.
The resulting range of $\ell= 1$ $g$-mode periods destabilized by the $\varepsilon$ mechanism spans from $150$ to $500\ $s, with radial order $k$ between $1$ and $4$. For $\ell= 2$ $g$ modes, the location in the  $\log g-T_{\rm eff}$ diagram is similar, but less modes become excited in this case. 

Up to our knowledge, no pulsating low-mass  He-core WD on its early-cooling branch with $M_{\star} \gtrsim 0.18- 0.20 \ M_{\sun}$ has been detected lying in the region of instability predicted in this work. However, there are some possible candidate stars, as illustrated by Fig. \ref{instability_HR}. 
The eventual detection of $g$ mode pulsations in low-mass  He-core stars populating this new instability domain would confirm the theoretically predicted existence of the $\varepsilon$ mechanism as an agent able to destabilize $g$-mode periods. Since the magnitudes of the rate of period change of $g$ modes for models evolving in stages prior to the CNO flashes are significantly large \citep[particularly, in comparison to the WD and pre-WD stages,][]{2017A&A...600A..73C}, this quantity could be measured and, in that case, support the predicted occurrence of the CNO flashes, providing a first proof of the existence of these flashes, and thus confirming the predicted age dichotomy for low-mass He-core WDs \citep{2001MNRAS.323..471A,2013A&A...557A..19A}. Last but not least, the detection of these pulsations would also help in the classification of several stars with uncertain nature.


Although we are aware that detecting pulsations (and rates of change of periods) in this type of objects is not an easy task, we consider that searches for low-amplitude variability are worth doing. As already shown by the results from the Transiting Exoplanet Survey Satellite  \citep[{\it TESS},][]{2015JATIS...1a4003R} in the case of the two new pre-ELMVs reported by \cite{2020ApJ...888...49W}, and as the future space missions like {\it Plato} \citep{2018EPSC...12..969P} and {\it Cheops} \citep{Moya2018} will probably show, the continuous improvement in the quality of
the observations is likely to help in this regard.

\begin{acknowledgements}
We wish to thank our anonymous referee for the constructive
comments and suggestions that greatly improved the original version of
the paper.
Part of this work was supported by PICT-2017-0884 from
ANPCyT, PIP 112-200801-00940 grant from CONICET, grant G149 from University of La Plata. 
K.J.B.\ is supported by the National Science Foundation under Award AST-1903828.
This research has made use of NASA Astrophysics Data
System.
\end{acknowledgements}


\bibliographystyle{aa} 
\bibliography{paper-pul-vii} 

\end{document}